\documentclass[prd,amsmath,twocolumn,nofootinbib, eqsecnum]{revtex4-2}
\usepackage{amsthm}
\usepackage{amssymb}
\usepackage{graphicx}
\usepackage{color}
\usepackage{url}
\usepackage{enumerate}
\usepackage{color}
\usepackage{slashed}
\usepackage{graphicx}
\usepackage{natbib}
\usepackage{float}
\def\beq{\begin{align}}
\def\eeq{\end{align}}

\newif\ifnote
\notetrue
\long\def\??#1{\textbf{\color{red}---~??{}#1~---}}
\long\def\note??#1{\ifnote\??{#1}\fi}

\begin{document}

\title{Mass corrections to the DGLAP equations}

\author{A. Jamali Hafshejani}
\email{ahmad.jamalii86@gmail.com}

\author{A. Mirjalili}
\email{a.mirjalili@yazd.ac.ir (Correspondent author)}

\affiliation {Physics Department, Yazd University, P.O.Box, 89195-741, Yazd, Iran}

\begin{abstract}
We propose a mass-dependent MOM scheme to renormalize UV divergence of unpolarized PDFs at one-loop order. This approach which is based on a once subtracted dispersion relation does not need any regulator. The overall counterterms are obtained from the imaginary part of large transverse momentum region in loop integrals. 
The mass-dependent characteristic of the scheme yields to mass-dependent splitting functions for the DGLAP evolution equations.
While the flavor number is fixed at any renormalization scale, the decoupling theorem is automatically imposed by the mass-dependent splitting functions. The required symmetries are also automatically respected by our prescription.
\end{abstract}

\maketitle
\section{Introduction}

The Wilson operator product expansion (OPE) \cite{Wilson:1969zs,Brandt:1970kg} provides a systematic approach to the factorization of quantum chromodynamics (QCD) \cite{Collins:1989gx,Collins:1998rz}; namely, the separation of contributions from the long and short distances.
The universal parton distribution functions (PDFs) are factorized from the hard partonic scattering. The latter can be calculated within perturbative QCD, but the PDFs have to be determined in a global analysis using experimental data \cite{Accardi:2016ndt,Clark:2019krt} and the Dokshitzer-Gribov-Lipatov-Altarelli-Parisi (DGLAP) equations \cite{Dokshitzer:1977sg,Gribov:1972ri,Altarelli:1977zs} that provide the scale evolution of the PDFs.
In an alternative approach, the programs for the OPE can be performed in terms of matrix elements of gauge invariant non-local operators giving explicit definition of bare PDFs \cite{Collins:1981uw}. In this fashion, the bare PDFs are ultraviolet (UV) divergent in perturbation theory, and require renormalization. 
The effective UV cutoff for the matrix elements is referred to as the \textit{factorization scale}, separating the short-distance from the PDFs.
Consequently, the DGLAP equations appear as renormalization group equations (RGEs) for PDFs in which the kernels (Altarelli-Parisy splitting functions \cite{Altarelli:1977zs}) play the roles of anomalous dimensions of the PDFs. Following this approach, one comes up with different DGLAP equations in different renormalization schemes. 

Conventional minimal subtraction (MS) schemes, as of mass-independent, could be good choices if we were not to deal with heavy quarks.
Working within MS renormalization schemes, one encounters large logarithms of $m_h/\mu$ when $\mu\ll m_h$, where $m_h$ is a heavy quark mass and $\mu$ is the renormalization scale. To avoid this, one would follow an effective theory to decouple heavy quarks and absorb their mass effects in the renormalized coupling constant. Alternative approach is to take a composite scheme, using a different prescription to renormalize diagrams containing heavy quarks (proposed by Collins-Wilczek-Zee (CWZ) scheme \cite{Collins:1978wz}).
However, requirement to a definition of \textit{heavy quarks} with respect to $\mu$ is inevitable. Unfortunately, heavy quark masses spread over a wide range of scales from a few to hundreds of GeVs.
Consequently, we have to work with a series of subschemes each of which considers different number of heavy (decoupled) and light (active) quarks. Then, some definite thresholds should be considered to switch between the subschemes. In addition, matching conditions at the thresholds should be imposed.
A number of such variable flavor number (VFN) schemes have been proposed for DIS structure functions, including ACOT \cite{Aivazis:1993kh,Aivazis:1993pi} S-ACOT \cite{Collins:1997sr,Kramer:2000hn}, ACOT-$\chi$ \cite{Tung:2001mv}, TR and TR' \cite{Thorne:1997ga,Thorne:2006qt}, and FONLL \cite{Buza:1996wv,Cacciari:1998it,Forte:2010ta}, etc. 
For a review on heavy quark mass effects in DIS and global analyses see \cite{Thorne:2008xf}.

On the other hand, as discussed in \cite{deOliveira:2013tya}, the critical issue of these schemes is that changes of the number of active quarks would lead to jumps in the splitting functions and the renormalized coupling constant.
As a solution, a new scheme is introduced in \cite{deOliveira:2013iya,deOliveira:2013tya,deOliveira:2013maa}.
Providing mass-dependent splitting functions, this scheme automatically results in a smooth behavior of PDFs, hard coefficient functions, and coupling constant. 
Although the idea of mass-dependent splitting functions had been already proposed in \cite{Martin:1996eva}, the heavy quark part was included only at large enough scale above the heavy quark mass. This approach gives irregularities at the heavy quark thresholds, and as shown in \cite{Olness:1997yn}, the approach followed in \cite{Martin:1996eva} results in the same physical cross sections as what are obtained by ACOT scheme. 

What we propose in this paper is a fixed-flavor-number mass-dependent MOM scheme based upon analytic structure of the Feynman diagrams.
Although mass-dependent MOM schemes decouple heavy quarks automatically and smoothly, they mostly suffer from complicated calculations and violation of symmetries of the theory in question. Therefore, defining a suitable mass-dependent MOM scheme requires careful considerations.
Our scheme, on one hand, provides automatically decoupling theorem, and on the other hand, it is lacking in the mentioned issues of a typical mass-dependent MOM scheme. 
All the required symmetries for PDFs, conservation of total momentum and flavor numbers, are automatically respected by our approach. 
Additionally, the possibility of extension to higher orders is supported by provided computational simplifications.
In our scheme, overall counterterms are extracted from off-shell Green functions of parton operators at large transverse momentum. This is carried out by once subtracted dispersion relation (DR) due to the analytic properties of the Feynman diagrams. 

The outcomes are mass-dependent splitting functions followed by mass corrections to the DGLAP equations.
The MOM characteristic of our scheme implies that decoupling theorem is imposed automatically and smoothly by mass correction terms in the splitting functions, while the flavor number is fixed.
This scheme also yields following computational simplifications: One is to deal with finite cut diagrams at just large transverse momenta region (full evaluation is not needed). Therefore, the requirement of any regulator vanishes.

This article is organized as following: In the section \ref{Definitions} basic and required definitions of the unpolarized PDFs and the DGLAP equations is briefly reviewed. 
Then, we introduce our scheme in the section \ref{Scheme}. Some detailed calculations of renormalized Green functions are also given to clarify the prescription.
In the section \ref{Splittings} mass-dependent splitting functions are presented and their following features are discussed.
The conclusion is given in the \ref{Conclusion}.
\section{Definitions}\label{Definitions}
We begin with a brief review on the theoretical framework of unpolarized PDFs and their renormalization.
Bare PDFs can be defined as diagonal matrix elements of Fourier transformed of gauge-invariant bi-local light-ray operators \cite{Collins:1981uw}
\begin{align} \label{Oi}
  O_{q_i}(k^+)\equiv\int &\frac{\text{d}w^-}{4\pi}\text{e}^{-\text{i}k^+w^-}
  \\\nonumber
  &\times :\overline{\psi}_{(0)}^i(nw^-)\gamma^+U_\text{F}(uw^-,0)\psi_{(0)}^i(0):
\end{align}
and
\begin{align}\label{Og}
  O_g(k^+)\equiv &\int\frac{\text{d}w^-}{-2\pi k^+}\text{e}^{-\text{i}k^+w^-}
  \\ \nonumber
  &\times:G^{+j}_{(0)}(nw^-)U_\text{A}(uw^-,0)G^{+}_{(0)j}(0):\;,
\end{align}
where $\psi_{(0)}^i$ is the bare quark field of flavor $i$, and $G^{\mu\nu}_{(0)}$ is the bare field strength of gluon. Sum over $j=1$ and $2$ is understood in Eq. Eq. (\ref{Og}).
Vector $u^{\mu}\equiv g^\mu_-$ points in minus light-cone direction. We use the following definition for light-cone coordinates of a given vector $v$:
\begin{equation}
  v^{\pm}\equiv(v^0\pm v^3)/\sqrt{2}\;.
\end{equation}
Operator $U(uw^-,0)$ is a Wilson line path ordered along $u$ \cite{Collins:1981uw}:
\begin{equation}
U(uw^-,0) = \mathcal{P} \exp \left[ - \text{i}g_{(0)}\int_0^{w^-} \text{d}{y^ - }A_{(0)}^{a,+}(uy^-)t_a \right],
\end{equation}
where $A_{(0)}^{a,\mu}$ is the bare gauge filed.
Indices F and A in Eq. Eq. (\ref{Oi}) and Eq. Eq. (\ref{Og}) refer to ``fundamental'' and ``adjoint'' representation of SU(3) group generators $t_a$.

Consider a hadron $h$ moving in the $z$ direction with four momentum $P$ expressed by $|P\rangle$ state. The probability of finding a parton of flavor $i$ with light-cone momentum fraction $\xi\equiv k^+/P^+$ within the hadron $h$, as bare PDF, is given by the matrix element as it follows
\begin{align}\label{f0}
f_{{(0)}\,i/h}&(\xi) \equiv \langle P|O_{i}(\xi P^+)|P\rangle_{\text{con}}\;,
\end{align}
where subscribe ``con'' implies that just connected diagrams should be taken into account.

The PDFs in Eq. Eq. (\ref{f0}) have an UV divergence arisen from not only the bare field operators but also the operator product in Eq. (\ref{Oi}) and Eq. (\ref{Og}). This requires us to apply a renormalization procedure using some renormalization factors and introducing a renormalization scale $\mu$ as
\begin{equation} \label{fR}
  f_{\text{(R)}i/h}(\xi,\mu)=Z_{ij}(\xi,\mu)\otimes  f_{\text{(0)}j/h}(\xi),
\end{equation}
where $\otimes$ is the conventional convolution in $\xi$ parameter:
\begin{equation}
  g(\xi,...)\otimes h(\xi,...)\equiv\int_\xi ^1 {\frac{{{\rm{d}}x}}{x}g\left(
{x,...}
\right)h\left( {\xi /x,...} \right)}.
\end{equation}
At leading order accuracy, renormalization factors should be 
\begin{equation}\label{Z0}
  Z^{[0]}_{ij}(\xi)=\delta_{ij}\delta(1-\xi)
\end{equation}
due to the fact that the renormalized and bare PDFs are the same at leading order. 
Notice the sum over $j$ in Eq. (\ref{fR}) runs over gluon as well as all flavors of quark and antiquark.
Renormalization group equations for the renormalized PDFs imply the following DGLAP evolution equations,
\begin{equation}
  \mu^2 \frac{\text{d}\mu^2}{\mu^2}f_{\text{(R)}i/h}(\xi,\mu)=P_{ij}(\xi,\mu)\otimes  f_{\text{(R)}j/h}(\xi).
\end{equation}
The kernels of the DGLAP equations are the anomalous dimension of renormalized PDFs and are referred as splitting functions. They can be written in matrix form as
\begin{equation}
  P(\mu)=\mu^2 \frac{\text{d}\mu^2}{\mu^2}\ln Z(\mu)\;.
\end{equation}
Therefore splitting functions are actually scheme dependent.

Since the renormalization factors are independent of the hadron state, it would be convenient to replace the hadron state with a parton state, in the given definition by Eq. (\ref{f0}). This gives parton in parton distributions which can be carried out in the framework of perturbative QCD by means of the Lehmann-Symanzik-Zimmermann (LSZ) redundant formula \cite{Lehmann:1954rq}. Therefore, the bare unpolarized distribution of parton $i$ in quark $j$ reads
\begin{equation}
  f_{(0)i/j}(\xi)=\frac{1}{6}\lim_{p^2 \to \overline{m}_j^2} \delta_{ab}(\slashed{p}+\overline{m}_j)_{\alpha \beta}\Gamma_{(0)ij}^{ab,\alpha \beta}(\xi p^+,p)\;.
\end{equation}
The factor $1/6$ is a consequence of average over spins and colors of the target parton $j$. 
$\Gamma_{(0)ij}$ is a bare amputated Green function of parton operator $O_i$, defined in Eq. (\ref{Oi}) or Eq. (\ref{Og}), accompanied by two field operators $\psi_{(0)}^j$ and $\overline{\psi}_{(0)}^{j}$ with off-shell external momenta $p$ and $-p$:
\begin{equation}\label{fij}
  \Gamma_{(0)ij}^{ab,\alpha \beta}(k^+,p)\equiv \langle 0|\mathcal{T} O_{i}(k^+)\psi^{a,\alpha}_{(0)j}(p)\overline{\psi}^{b,\beta}_{(0)j}(-p)|0\rangle_{\text{amp}}\;,
\end{equation}
where ``amp''-subscribe refers to ``amputated'' Green function and $\overline{m}$ is physical mass. Note that Fourier transformations are defined outside the time order operator.
We work with bare fields (instead of physical ones) for which loops on external lines should be taken into account. Note that the Green function in Eq. (\ref{fij}) is time ordered which gives the same result as of fixed ordered as long as we work in a covariant gauge \cite{Collins:2011zzd}.
Coordinate system is chosen such that the external momentum has no transverse component,
\begin{equation}
  p^\mu=g^\mu_+ p^+ + g^\mu_- p^2/2p^+\;.
\end{equation}
In the same manner, the bare unpolarized distribution of a parton $i$ in gloun can be written in terms of an amputated Green function as 
\begin{equation}\label{fig}
   f_{(0)i/g}(\xi)=\frac{1}{16}\lim_{p^2 \to \overline{m}_j^2} {\delta_{ab}d_{\mu\nu}(p)}\Gamma_{(0)gj}^{ab,\mu \nu}(\xi p^+,p)\;,
\end{equation}
where the amputated Green function reads
\begin{equation}\label{GR1qg}
  \Gamma_{(0)ig}^{ab,\mu\nu}(k^+,p)\equiv \langle 0|\mathcal{T} O_{i}(k^+)A_{(0)}^{a,\mu}(p)A^{b,\nu}_{(0)}(-p)|0\rangle_{\text{amp}}\;.
\end{equation}
The sum over physical polarizations in Eq. (\ref{fig}) is done by following tensor
\begin{equation}\label{d}
  d_{\mu\nu}(p) \equiv -g_{\mu\nu}+\frac{p_\mu g^+_\nu + p_\nu g^+_\mu}{p^+}\;.
\end{equation}
and the factor $1/16$ resulted from average over physical polarizations and colors of the initial gluon.

From Eq. (\ref{fij}) and Eq. (\ref{fig}), renormalized parton in parton distributions can be defined by replacing the bare amputated Green functions with their renormalized version.
On the other hand, regarding the standard definition, given by Eq. (\ref{fR}), one can consider the same renormalization relation for the Green functions as
\begin{equation} \label{GaR}
  \Gamma_{\text{(R)}ij}(\xi p^+,p;\mu)= Z_{ik}(\xi,\mu)\otimes \Gamma_{(0) kj}(\xi p^+,p)+\dots ,
\end{equation}
where the dots indicate contributions of Green functions of possible gauge un-invariant (but BRST invariant) operators mixing with the parton operators in the renormalization process \cite{Collins:1984xc}. These unphysical contributions, however, are not our concern as long as they are killed by the equations of motion imposing by LSZ redundant formula.
Therefore, these unphysical Green functions do not contribute to the PDFs renormalization.
In general, renormalization factors may have tensor and spinor indices contracting with Green functions, but we assume that in an appropriate scheme they appear to be scalar coefficients. This is justified in our scheme in the section \ref{Scheme} for the cases in which the target is a quark. According to this assumption, renormalization factors in Eq. (\ref{fR}) and Eq. (\ref{GaR}) turn to be equal.

\section{Our proposed renormalization scheme}\label{Scheme}

The PDFs are supposed to contain information about large distance properties of hadrons, therefore we propose a renormalization prescription to remove all short distance behaviors of the bare PDFs.
To describe our scheme, let us first consider a dispersion relation (DR).
Given the imaginary part of an analytic function $f(x)$, one can write a DR as:
\begin{equation}\label{DR}
  f(x)=\int^\infty_{x_\text{min}}\frac{\text{d}s}{s-x-\rm{i}\epsilon} \;\frac{\text{Im} f(s)}{\pi}.
\end{equation}
If the integral is logarithmically divergent, a once subtracted DR can be performed to obtain a finite result
\begin{equation}\label{1DR}
  f(x)-f(c)=\int^\infty_{x_\text{min}}\text{d}s\frac{x-c}{(s-x-\text{i}\epsilon)(s-c)} \frac{\text{Im} f(s)}{\pi},
\end{equation}
where $c<x_\text{min}$ to avoid a singularity. 
Consider we are to impose the following renormalization condition on a given diagram, $\Gamma^{(\gamma)}$:
\begin{equation}
  \Gamma^{(\gamma)}_{(\rm{R})}(p^2;\mu)=\Gamma^{(\gamma)}(p^2)-\Gamma_{\rm{UV}}^{(\gamma)}(p^2=-\mu^2),
\end{equation}
according which just the UV-divergent piece is subtracted off.
This can be carried out by use of a once subtracted DR Eq. (\ref{1DR}) in which the total imaginary part is replaced by a particular piece of the imaginary part generating just the UV divergence:
\begin{align} \label{GR} 
  &\Gamma^{(\gamma)}_{(\rm{R})}(p^2;\mu)=
  \\ \nonumber
    &\int^\infty_{p^2_\text{min}}\text{d}s\frac{s+\mu^2}{(s-p^2-\text{i}\epsilon)(s+\mu^2)} \frac{\text{UIm} \Gamma^{(\gamma)}(s)}{\pi}+\Gamma^{(\gamma)}_{\rm{IR}}(p^2).
\end{align}
Here ``UIm'' refers to as ``the particular term of the imaginary part that generates the UV divergence''. 
Notice, in Eq. (\ref{GR}), just the UV divergence is subtracted and the UV finite part, indicated by $\Gamma^{(\gamma)}_{\rm{IR}}$, is left unchanged.
In this renormalization scheme, one is supposed to find just the UIm of a diagram instead of its total imaginary part. This can be simply done by using cut diagrams at UV region of the loop integral, i.e., the limit of infinite transverse momenta. This can be also verified by power counting of $p^2$.
Note that a sum over all possible legitimate cut versions of a given diagram amounts to its total imaginary part,
\begin{equation}
  \text{Im}\;\Gamma^{(\gamma)}=\frac{1}{2}\sum_{\text{cut}}\Gamma^{(\gamma)}_{\text{cut}}\;.
\end{equation}
Requiring just UIm of the diagrams, our prescription is followed by two simplifications:
\begin{enumerate}
  \item Just UV limit of the cut diagrams should be calculated instead of their total value.
  \item  There is no need for a regulator since UIms are finite.
\end{enumerate}
What we do in the rest of this section is to calculate UIm of the amputated Green functions to renormalize them by Eq. (\ref{GR}).

\begin{figure}
{\includegraphics[scale=.5]{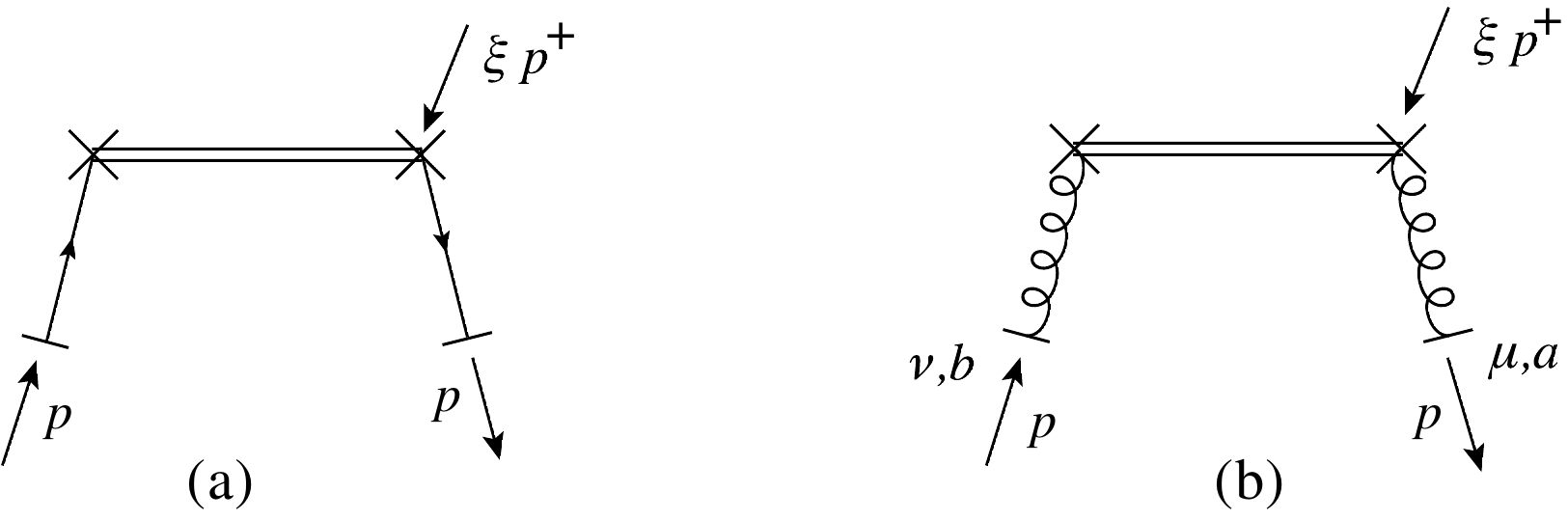}}
\caption{Cut diagrams of (a) quark in quark and (b) gluon in gluon amputated Green function at leading order. \label{LO}}
\end{figure}

At leading order, we have two non-vanishing Green functions, quark in quark type, Fig.\ref{LO}(a), and gluon in gluon type, Fig.\ref{LO}(b), as in following
\begin{equation} \label{f0ij}
  \Gamma^{[0]\alpha\beta}_{ij,ab}(\xi k^+,p)=\frac{\gamma^+_{\alpha\beta}}{2p^+} \sigma(p^+)\delta_{ab}\delta_{ij}\delta\left(1-\xi\right)\;,
\end{equation}
\begin{equation} \label{f0gg}
  \Gamma^{[0]\mu\nu}_{gg\;ab}\left(\xi k^+,p\right)=g^{\mu i} g^{\nu}_i \delta_{ab} \sigma(p^+)\left[\delta\left(1-\xi\right)-\delta(1+\xi)\right]\;.
\end{equation}
From now on, we consider particle in particle distributions, i.e, for which $p^+$ and $k^+>0$, and suppress flavor and color indices (as long as they are just Kronecker deltas). Spinor indices, which are explicit from $\gamma$-matrices, are also suppressed. It is convenient to indicate the dependence of $k^+$ through the $\xi=k^+/p^+$ ratio.

Substituting the leading order Green functions and renormalization factors, Eq. (\ref{f0ij}), Eq. (\ref{f0gg}), and Eq. (\ref{Z0}), in Eq. (\ref{GaR}), one gets the following expressions at one-loop order:
\begin{equation}
  \Gamma^{[1]}_{(\text{R})qq}\left(\xi,p;\mu\right)=\Gamma^{[1]}_{(0)qq}\left(\xi,p\right)+\frac{\gamma^+}{2p^+}Z^{[1]}_{qq}\left(\xi,\mu\right)\;,
\end{equation}
\begin{align}
  \Gamma^{[1]}_{(\text{R})gq}\left(\xi,p;\mu\right)&=\Gamma^{[1]}_{(0)gq}\left(\xi,p\right)+\frac{\gamma^+}{2p^+}Z^{[1]}_{gq}\left(\xi,\mu\right)\;,
\end{align}
\begin{align}
  \Gamma^{[1]\mu\nu}_{(\text{R})qg}\left(\xi,p;\mu\right)=&\Gamma^{[1]\mu\nu}_{(0)qg}\left(\xi,p\right)+g^{\mu i} g^{\nu}_i Z^{[1]}_{qg}\left(\xi,\mu\right)\;.
  \end{align}

\begin{figure}
{\includegraphics[scale=.50]{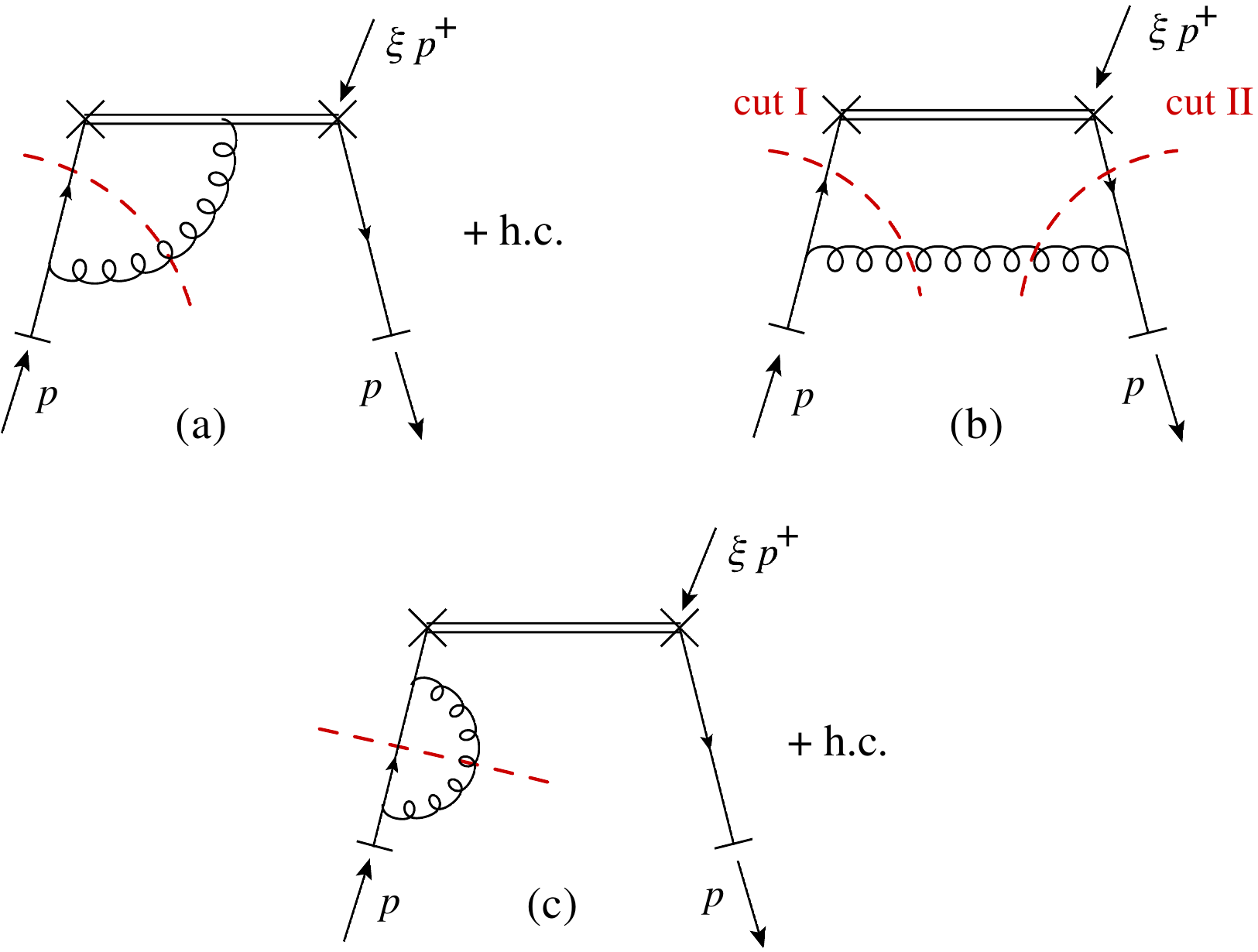}}
\caption{ Cut diagrams of quark in quark amputated Green function at one-loop order. \label{qq}}
\end{figure}

We first start with the most important case of quark in quark. Detailed calculations are represented in order to clarify our prescription.
Consider cut diagrams of quark in quark PDF in Fig.\ref{qq}. The total value of diagrams (a) and (a$^\dagger$), indicated by ``h.c.'' in Fig.\ref{qq}(a), are exactly UIm, which gives
\begin{equation}
 \frac{1}{2}\Gamma^{(\rm{a}+\rm{a}^\dagger)}_{\text{cut}}\left(\xi,p\right)
=\frac{g^2}{16\pi}C_\text{F}\frac{\gamma^+}{p^+}  \left[\frac{2\xi}{1-\xi}\theta(p^2-m^2/\xi)\right]_+,
\end{equation}
where a plus distribution is defined as
\begin{equation}
  \left[f(\xi)\right]_+\equiv f(\xi)-\theta(0<\xi<1)\delta(1-\xi)\int_0^1\text{d}\alpha\;f(\alpha)\;,
\end{equation}
and $C_\text{F}$ is the value of the quadratic Casimir operator of SU(3) group in the fundamental representation.

On the other hand, for the case of the cut diagram in Fig.\ref{qq}(b) we need to extract UIm from the total imaginary part which is given by
\begin{align} \label{Gb} \nonumber
& \frac{1}{2}\left(\Gamma^{(\text{b})}_\text{cutI}+ \Gamma^{(\text{b})}_\text{cutII}\right)\left(\xi ,p\right) =\frac{-g^2}{16\pi}C_\text{F} (1-\xi)\theta(0<\xi<1)
 \\
 &\times\int^\infty _0 \text{d}{\bf{q^2_T}}\frac{{\bf{q^2_T}}\frac{\gamma^+}{p^+}+2\xi^2p^+\gamma^- -4m\xi}{{\bf{q^2_T}}+M(\xi,p^2)}\delta({\bf{q^2_T}}+M(\xi,p^2)),
\end{align}
where $M(\xi,p^2)\equiv (1-\xi)(m^2-\xi p^2)$.
As mentioned above, UIm can be derived from UV region of the integral, i.e., ${\bf{q^2_T}}\to \infty$. Therefore we have
\begin{align} \nonumber
  \text{UIm}\;\Gamma^{(\text{b})}\left(\xi,p\right)=&-\frac{g^2}{16\pi}C_\text{F} \frac{\gamma^+}{p^+} (1-\xi)
  \\
  &\times\theta(0<\xi<1)\theta\left(p^2-m^2/\xi\right)\;.
\end{align}
The contribution of the diagrams (c) and (c$^\dagger$) amount to
\begin{align}
&\frac{1}{2}\Gamma^{(\rm{c}+\rm{c}^\dagger)}_\text{cut}(\xi,p)=\frac{g^2}{16\pi}\int_0^1 \text{d}\alpha (1-\alpha) 
\\ \nonumber
&\times\int_0^{+\infty}\text{d}{\bf{q^2_T}} \delta\left({\bf{q^2_T}}+M(\alpha,p^2)\right)\frac{\left({\bf{q^2_T}}-m^2\right)\frac{\gamma^+}{p^+}+\alpha(\alpha-2)m^2}{{\bf{q^2_T}}-(1-\alpha)^2m^2}\;,
\end{align}
which the UV limit of the integral gives the associated UIm as
\begin{align}
  \text{UIm}\Gamma^{(\rm{c}+\rm{c}^\dagger)}(\xi,p)=\frac{g^2}{16\pi}C_\text{F}\frac{\gamma^+}{p^+}\int_0^1 \text{d}\alpha (1-\alpha)\theta\left(p^2-m^2/\xi\right).
\end{align}
Therefore, by substituting UIm in Eq. (\ref{GR}), the renormalaized Green function of quark in quark amounts to 
\begin{align} \label{GRqq}
  &{\Gamma}^{[1]}_{(\text{R})qq}(\xi,p;\mu)=
  \\ \nonumber
  -&\frac{g^2}{16\pi^2}C_\text{F}\frac{\gamma^+}{p^+}\left[ \frac{1+\xi^2}{1-\xi}\int^\infty_{\frac{m^2}{\xi}} \frac{\text{d}s}{s-p^2}\frac{\mu^2+p^2}{\mu^2+s}\right]_++\Gamma_\text{IR}(\xi,p)\;.
\end{align}
Notice the pure plus distribution form of the counterterm resulting in conservation of each flavor number.
The renormalized Green function Eq. (\ref{GRqq}) is identical to subtracting the logarithm part of the bare one at renormalization point $-\mu^2$ as well as pole part of a regulator. 
Therefor, oppose to a MS scheme, we do not have large logarithm of $m/\mu$ here. To avoid these large logarithms in a MS scheme, either one should get heavy quarks decoupled at low renormalization scales, $\mu\ll m$, or use a mass-dependent scheme for diagrams involving heavy quarks (like CWZ scheme \cite{Collins:1978wz}).

\begin{figure}
{\includegraphics[scale=.5]{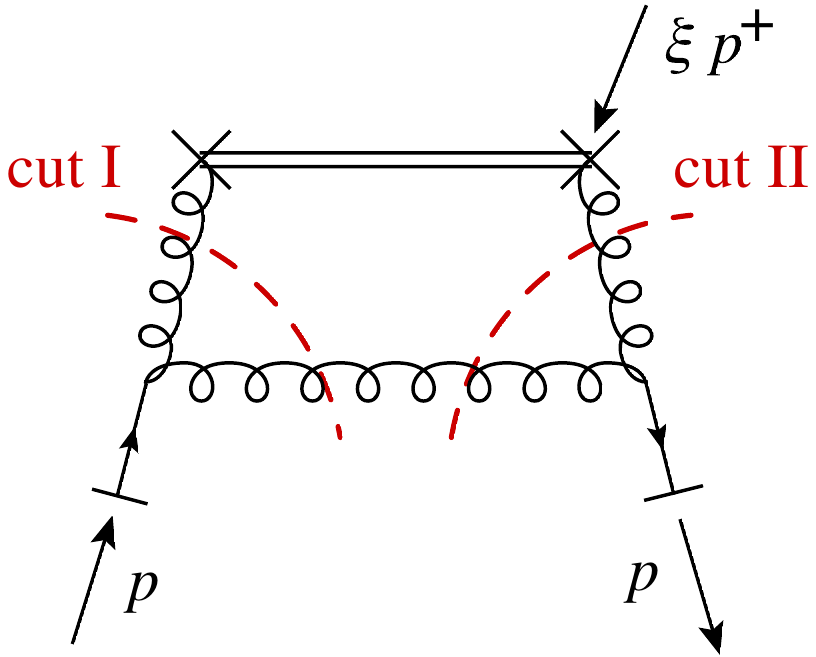}}
\caption{ Cut diagrams of gluon in quark amputated Green function at one-loop order. \label{gq} }
\end{figure}

The same approach is applied to the other cases.
There is one diagram with two cuts for the Green function of gluon in quark, depicted in Fig.\ref{gq}. Evaluated at large transverse momenta, the cut diagrams give
\begin{align}
  \text{UIm}&\Gamma^{[1]}_{(0)gq}(\xi,p;\mu)=
  \\ \nonumber
  -&\frac{g^2}{8\pi} C_{\text{F}}\frac{\gamma^+}{p^+}\frac{ 2-2\xi+\xi^2}{\xi}\;\theta\left(p^2-\frac{m^2}{\xi (1-\xi)} \right).
\end{align}
Notice the transformation $\xi \to 1-\xi$ under which the counterterm of gluon in quark Green function transforms to counterterm of quark in quark Green function. This manifests the conservation of total momentum.

\begin{figure}
{\includegraphics[scale=.50]{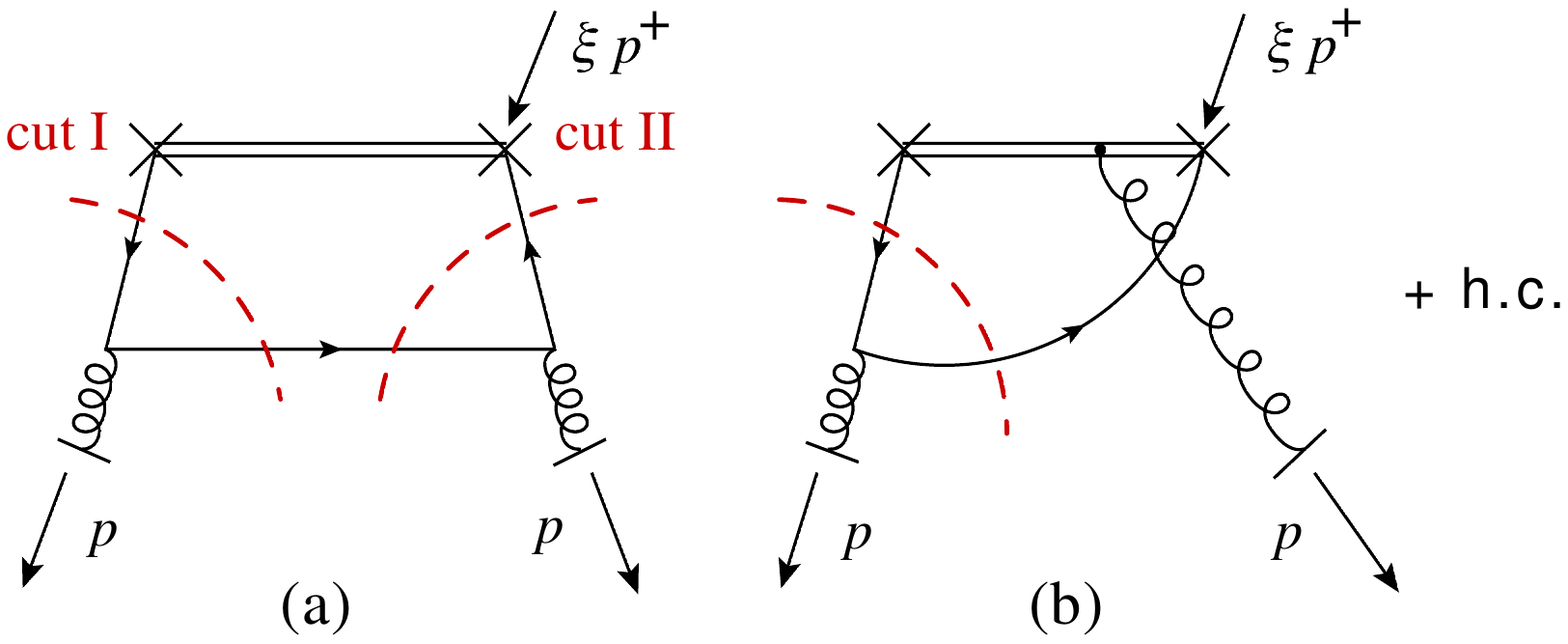}}
\caption{Cut diagrams of quark in gluon amputated Green function at one-loop order.\label{qg}}
\end{figure}

In general we have three cut diagrams associated with the case of quark in gluon, illustrated in Fig.\ref{qg}.
To get rid of (b) we can project the Green function by $d_{\mu\nu}(p)$, defined in Eq. (\ref{d}). By so doing, we sum over physical polarizations which results in extracting renormalization factors for unpolarized target gluon:
\begin{equation}
  Z^{[1]}_{qg}(\xi,\mu)=\frac{1}{2}\left[ d_{\mu\nu}(p)\Gamma^{(a)\mu\nu}_{qg}(\xi,p)-d_{\mu\nu}(p)\Gamma^{(a)\mu\nu}_{(\text{R})qg}(\xi,p;\mu) \right].
\end{equation}
Therefore, to find the renormalization factor we just need to obtain UIm of the projected graph (a), which would be
\begin{align}
  \text{UIm}& \left[d_{\mu\nu}(p)\Gamma^{(a)\mu\nu}_{qg}(\xi,p)\right]
  \\ \nonumber
  &=-\frac{g^2}{8\pi}T_{\text{R}}\left( 2\xi^2-2\xi+1\right)\theta\left(p^2-\frac{m^2}{\xi (1-\xi)} \right)\;,
\end{align}
where $T_\text{R}$ is conventional notation for the normalization of the SU(3) group generators.

\section{Mass-dependent splitting functions}\label{Splittings}

\begin{figure}[t]
{\includegraphics[scale=.82]{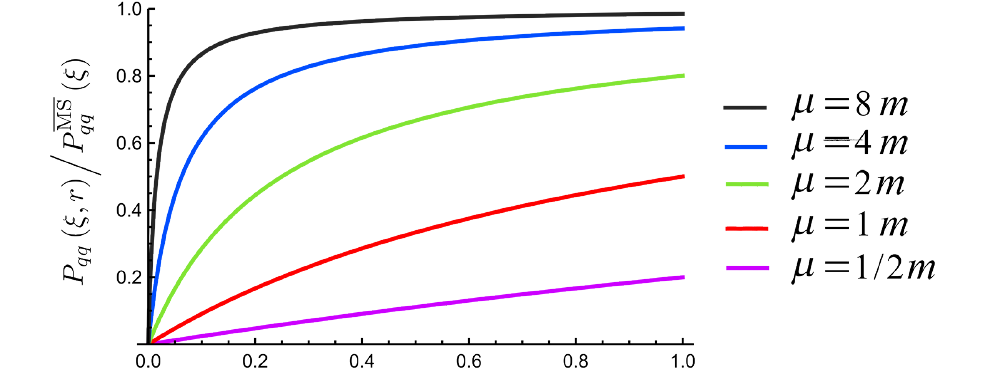}}
\caption{The effect of the mass-dependent coefficient on the splitting function $P_{qq}$. \label{Pqq_plot}}
\end{figure}
\begin{figure}
{\includegraphics[scale=.6]{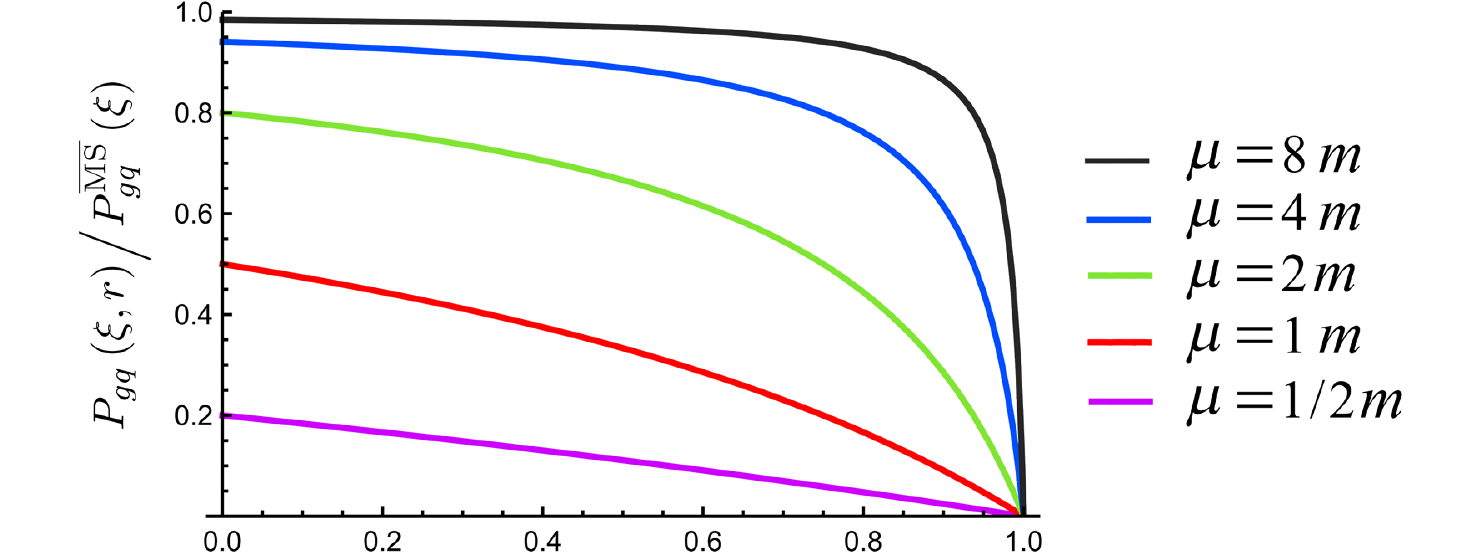}}
\caption{ The effect of the mass-dependent coefficient on the splitting function $P_{gq}$. \label{Pgq_plot}}
\end{figure}
\begin{figure}
{\includegraphics[scale=.6]{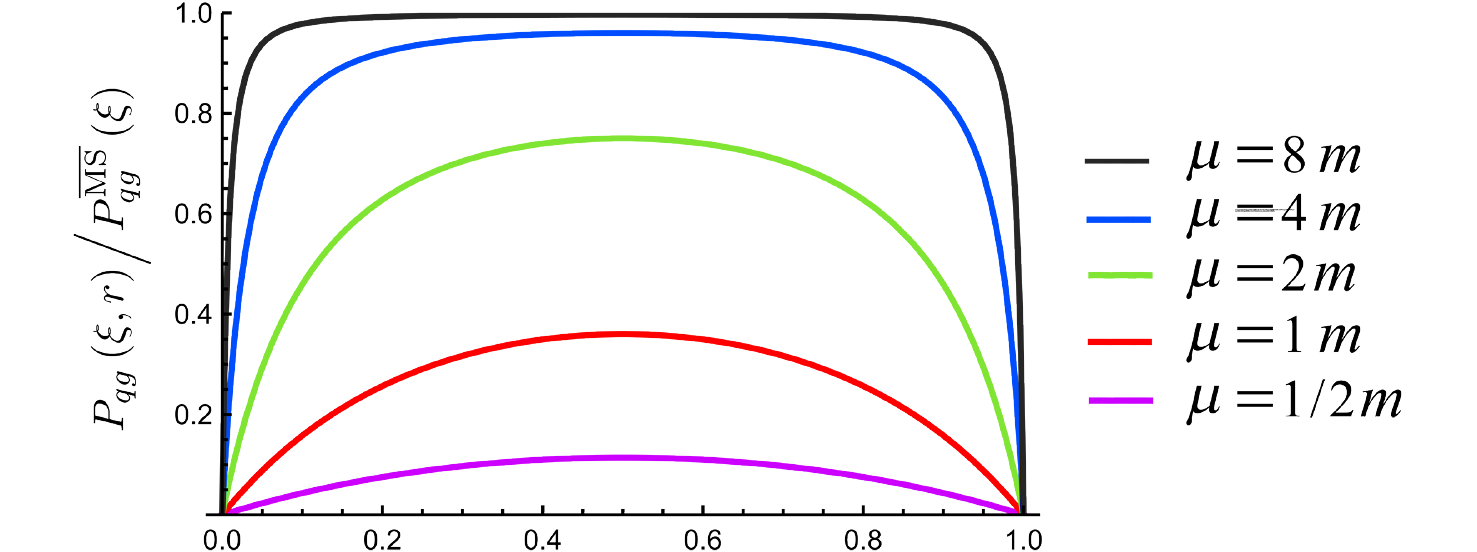}}
\caption{ The effect of the mass-dependent coefficient on the splitting function $P_{qg}$.\label{Pqg_plot}}
\end{figure}

In this section we represent mass-dependent splitting functions obtained by the scheme introduced in this paper. In general, resulted splitting functions are proportional to the conventional ones multiplied by a mass-dependent coefficient. We show that these coefficients automatically impose a smooth decoupling on the DGLAP equations.

As mentioned, splitting functions can be obtained as the anomalous dimension of the parton in parton PDFs, which at one-loop order gives
\begin{equation}\label{P1}
  P_{ij}(\xi,\mu)=-\mu^2 \frac{\rm{d}}{\rm{d}\mu^2}\Gamma_{{(\rm{R})}ij}(\xi,p;\mu)
\end{equation}
Substituting the renormalized Green functions, e.g. Eq. (\ref{GRqq}), in Eq. (\ref{P1}), we obtain the mass-dependent splitting functions as below
\begin{equation}\label{Pqq}
  P^{[1]}_{qq}(\xi,r)=\frac{g^2}{8\pi^2}C_\text{F} \left[\frac{1+\xi^2}{1-\xi}\frac{\xi}{r+\xi}\right]_+\;,
\end{equation}
\begin{equation}\label{Pgq}
  P^{[1]}_{gq}(\xi,r)=\frac{g^2}{8\pi^2}C_\text{F} \left[\frac{1+(1-\xi)^2}{\xi}\frac{1-\xi}{r+1-\xi}\right],
\end{equation}
and
\begin{equation} \label{Pqg}
  P^{[1]}_{qg}(\xi,r)=\frac{g^2}{8\pi^2}T_\text{R}\left[ {1 - {{\left( {\frac{r}{{r + \xi \left( {1 - \xi }
\right)}}}
\right)}^2}} \right]\left[ {{\xi ^2} + {{\left( {1 - \xi } \right)}^2}} \right],
\end{equation}
where $r\equiv m^2/\mu^2$.
The conservation of quarks number for each flavor implies that $P_{qq}(\xi)$ should be in a form of a plus distribution, which is automatically satisfied in our scheme.
In addition, the conservation of the total momentum results in the symmetry
\begin{equation} \label{Pgq}
  P_{gq}(1-\xi)=P_{qq}(\xi),
\end{equation}
which is also automatically respected in our scheme.

Decoupling heavy quarks is automatically controlled by the mass-dependent terms in the splitting functions in Eq. (\ref{Pqq}), Eq. (\ref{Pgq}), and Eq. (\ref{Pqg}).
Plots depicted in Figs.\ref{Pqq_plot}, \ref{Pgq_plot}, and \ref{Pqg_plot} obviously reveal the fact that the mass-dependent splitting functions are forced to vanish at relatively low renormalization scales, $\mu \ll m$, and go to the conventional ones at relatively high renormalization scales, $\mu \gg m$.
This justifies the fact that one should sum over all 6 quark flavors in the DGLAP equations when using these mass-dependent splitting functions. Effects of the heavy quarks mass are automatically taken into accounts with respect to the renormalization scale, starting from the input scale $Q_0$ in the DGLAP equations.

\begin{figure}[b]
{\includegraphics[scale=.50]{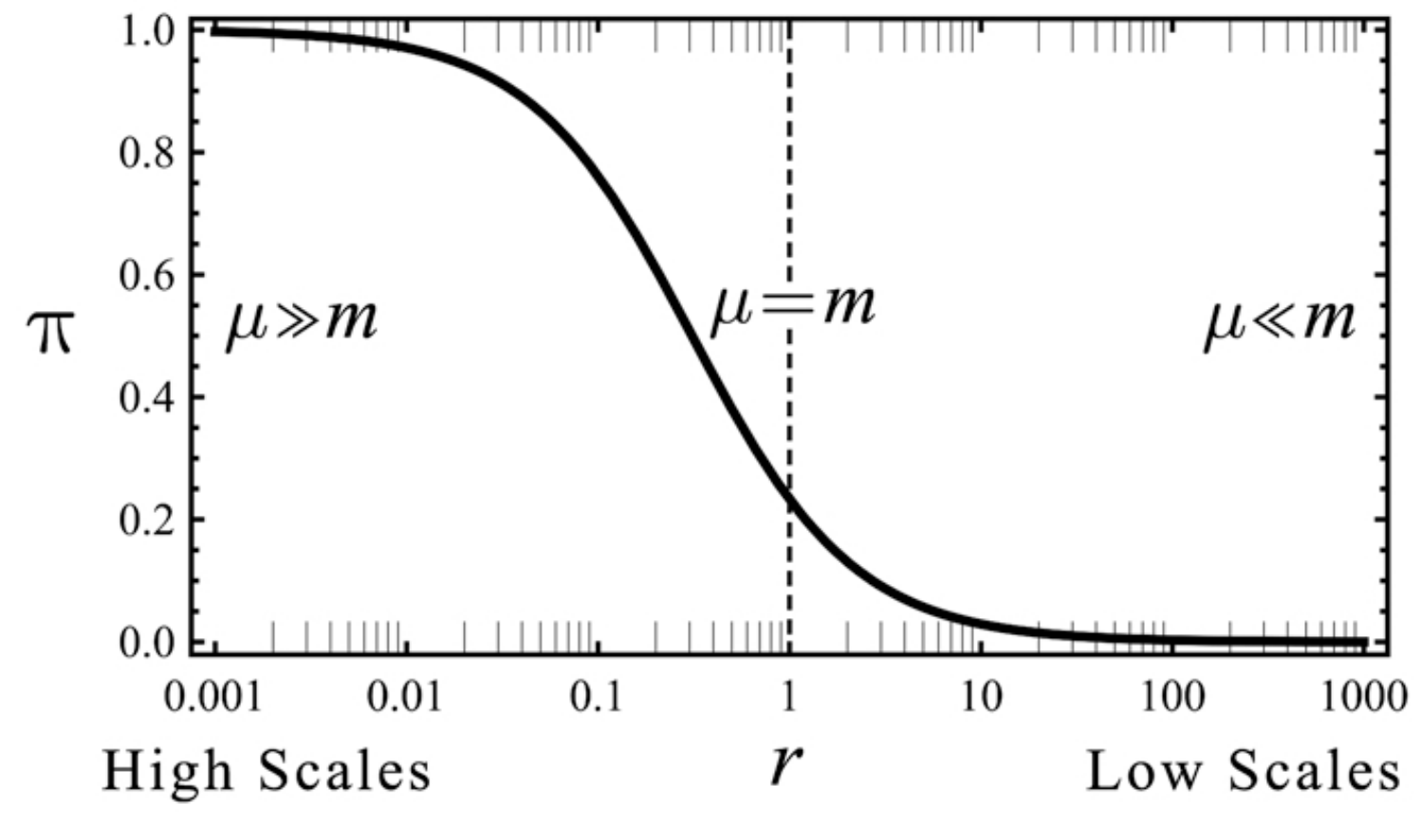}}
\caption{Function $\pi(r)$ which gives the contribution of each heavy quark loop to $P_{gg}$.
\label{pi_plot}}
\end{figure}

Mass corrections to the $P_{gg}$ splitting function, at one-loop order, should be performed to the coefficient of the $T_{R}$ in the delta function coefficient. This term is responsible for fermionic loops in the external gluon propagator.
The mass correction can be determined using sum-rule
\begin{equation}\label{SR2}
  \int_0^1 \text{d}\xi \;\xi  P_{gg}(\xi)+\sum_{i}^6 \int_0^1 \text{d}\xi\; \xi P_{q_{i}g}(\xi)=0\;,
\end{equation}
which is resulted from the conservation of the total momentum. Substituting the mass-dependent splitting function $P_{qg}$, Eq. (\ref{Pgq}), in the sum-rule given by Eq. (\ref{SR2}) we obtain the mass-correction to the conventional splitting function $P_{gg}$. That is a replacement of the flavor number $n_f$ in the conventional $P_{gg}$ with the summation
\begin{equation}\label{nf}
	\sum\limits_{i=1}^{6}\pi(r_i),
\end{equation}
where
\begin{equation} \label{pi}
   \pi \left( r \right) \equiv \frac{1}{1 + 4r}\left[ {1 + r - 6{r^2} + \frac{12r^3}{\sqrt {1 + 4r}}\ln\frac{\sqrt {1 + 4r}  + 1}{\sqrt {1 + 4r}  - 1}} \right].
\end{equation}
Note that the flavor number is fixed at $6$ in Eq. (\ref{nf}). Having a smooth behavior across the heavy quark thresholds, the function Eq. (\ref{pi}) is analogous with the step function $\theta(\mu-\mu_{\rm{threshold}})$ in VFN schemes. 
The plot in Fig.\ref{pi_plot} shows that $\pi(r)$ behaves like a very smooth step function, and smoothly adds the contribution of each heavy quark loop in gluon self-energy from low to high scales.
This contribution for each heavy flavor is completely decoupled at renormalization scales very lower that the quark mass.
For renormalization scales very higher than the quark mass, function $\pi(r)$ acts as if the quark were massless, as evidenced in Fig. Eq. (\ref{pi_plot}).

\section{Conclusion}\label{Conclusion}
In this work, we use a mass-dependent MOM scheme to renormalize unpolarized PDFs.
Working with this unconventional approach, one is to subtract large transverse momenta behavior (as well as UV pole part) of bare PDFs, which actually belongs to the hard region.
Once subtracted dispersion relation is applied to perform the subtraction using the imaginary part of the Feynman diagrams. We define a specific part of the total imaginary part being correspond to just UV divergence in the real part.
This approach is followed by computational simplifications and respecting symmetries.
The resulted splitting functions differ from the conventional ones by mass-dependent factors. We show that these corrections provide automatically decoupling heavy quarks. Therefore, the flavor number is fixed at 6 at any renormalization scale and decoupling is controlled by the mass-dependent splitting functions.

Some phenomenological works are needed to test the prescription. The PDF of gluon and light quarks at a low initial scale would be taken as the input for the DGLAP equations. In addition, the PDF for the heavy quarks at the initial scale should be set zero, and the summation in the DGLAP equation should runs over all 6 quark flavors.
The results will specifically show a tiny contribution for the heavy quarks distributions at renormalization scales below their masses. These contributions, however, would not be of large magnitude since the ratio of the mass-dependent splitting functions over the conventional ones are small even at scales $\mu\simeq m$, as can be seen in Figs. \ref{Pqq_plot}, \ref{Pgq_plot}, and \ref{Pqg_plot}.

\bibliography{Bibliography}

\end{document}